\documentclass[twoside]{article}
\usepackage[a4paper]{geometry}
\usepackage[latin1]{inputenc} 
\usepackage[T1]{fontenc} 
\usepackage{RR}
\usepackage{hyperref}
\usepackage{url}
\usepackage{xspace}
\newcommand{\cmo}{\textsc{cmo}\xspace}
\newcommand{\paul}{\textsc{paul}\xspace}
\newcommand{\dali}{\textsc{dali}\xspace}
\newcommand{\matras}{\textsc{mat\-ras}\xspace}
\newcommand{\dalix}{\textsc{dalix}\xspace}
\newcommand{\matrasx}{\textsc{mat\-rasx}\xspace}
\newcommand{\tmalign}{\textsc{tm-align}\xspace}
\newcommand{\dast}{\textsc{dast}\xspace}

\newcommand{\eg}{\emph{e.g.}\xspace}
\newcommand{\ie}{\emph{i.e.}\xspace}
\newcommand{\cf}{\emph{cf.}\xspace}

\RRNo{7874}
\RRdate{February 2012}
\RRauthor{
Inken Wohlers\thanks{To whom correspondence should be addressed; I.Wohlers@cwi.nl}\thanks[cwi]{Life Sciences, Centrum Wiskunde \& Informatica, Science Park 123, 1098 XG Amsterdam, the Netherlands} \and No\"{e}l Malod-Dognin\thanks{ABS, INRIA Sophia Antipolis - M\'{e}diterran\'{e}e} \and Rumen Andonov\thanks{GenScale, INRIA Rennes - Bretagne Atlantique and University of Rennes 1} \and \\Gunnar W. Klau\thanksref{cwi}}

\authorhead{Wohlers, Malod-Dognin, Andonov and Klau}
\RRtitle{CSA : Comparaison compr\'ehensible d'alignement de paires de structures de prot\'eines}
\RRetitle{CSA: Comprehensive comparison of pairwise protein structure alignments}
\titlehead{CSA: Comprehensive comparison of pairwise protein structure alignments}
\RRresume{
CSA est un serveur web pour la comparaison compréhensible des alignements de paires de protéines. Son moteur d'alignement exact calcule, pour les scores de similarité basés sur les distances inter-résidus CMO (Contact Map Overlap maximization), PAUL, DALI et MATRAS, soit des alignements optimaux (ayant le plus haut score), soit des alignements heuristiques avec une garantie de qualité.
Ces alignements, plus ceux uploadés par l'utilisateur, sont comparés en utilisant de nombreuses mesures de qualité et des visualisations intuitives.
CSA apporte un nouveau regard sur les relations structurales entre paires de protéines, et constitue un outil précieux pour l'étude des similarités structurales.
CSA est disponible sur \url{http://csa.project.cwi.nl}
}
\RRabstract{
CSA is a web server for the comprehensive comparison of pairwise protein structure alignments. Its exact alignment engine computes either optimal, top-scoring alignments or heuristic alignments with quality guarantee for the inter-residue distance based scorings of contact map overlap, PAUL, DALI and MATRAS. These and additional, uploaded alignments are compared using a number of quality measures and intuitive visualizations.
CSA brings new insight into the structural relationship of the protein pairs under investigation and is a valuable tool for studying structural similarities. It is available at \url{http://csa.project.cwi.nl}.
}
\RRmotcle{Alignement de structure protéiques, comparaison d'alignements, serveur web, algorithme exact, fonctions de score}
\RRkeyword{Protein structure alignments, alignment comparison, web server, exact algorithm, scoring function}
\RRprojets{ABS, GenScale}
\RCSophia 

\begin{document}
\makeRR   

\section{Introduction}

Protein structural alignment is a key method for answering many biological questions that involve the transfer of information from well-studied proteins to less well-known proteins. Since structures are more conserved during evolution than sequences, structural alignment allows for the most precise mapping of equivalent residues. It is notably important for (i) detecting and investigating structural motifs, functional sites, and common cores and (ii) measuring similarity between proteins and bringing them in evolutionary relationship, \eg, by classification. Numerous web servers are available that offer individual methods for computing structural alignments, \eg, \cite{Kawabata2003,Mosca2008,Margraf2009,Holm2010}.

Many structure-based scoring schemes have been proposed and there is no consensus which scoring is the best~\cite{Hasegawa2009}. Comparative studies find that scorings have individual strengths and weaknesses and that alignments produced by different methods can differ considerably~\cite{Mayr2007}. In the context of protein classification, there are first attempts to integrate information from alignments generated by different structural alignment methods, \eg, \cite{Camoglu2006,Barthel2007}.

Here, we present CSA (Comparative Structural Alignment), the first web server for comprehensive comparison of pairwise protein structure alignments at single residue level. CSA facilitates evaluating the agreement between and advantages of alignments that maximize different established scoring schemes. It offers the computation of alignments using the scoring schemes of \dali~\cite{Holm1993}, contact map overlap (\cmo) \cite{Godzik1994}, \matras~\cite{Kawabata2000}, and \paul~\cite{Wohlers2010}. CSA uses our own, exact algorithm \cite{Andonov2011,Wohlers2011} that can be used with any inter-residue distance based scoring scheme. We choose \cmo and \paul scoring since they are tailored to the algorithm and \dali and \matras scoring because they are established and their programs and web servers \cite{Kawabata2003,Holm2010} are widely used. CSA returns an optimal, \ie top-scoring alignment, if found within the time limit, or otherwise an alignment with a quality guarantee that specifies how much improvement is at most possible. We denote this by calling our program and its alignments \dalix and \matrasx, in which \textsc{x} indicates exact. 

Optimality comes at the prize of higher running time, but is especially important when comparing alignments. A top-scoring, but biologically implausible, alignment implies that the scoring scheme used is inadequate to detect the given structural relationship and a different scoring might be more advisable. In the case of pairwise structural alignment, in which primarily residue correspondences are of interest, and only secondarily the obtained similarity score, comparing alignments optimized with respect to different criteria thus brings additional insight. 

In CSA, computed or uploaded alignments can be explored in terms of many inter-residue distance-, RMSD- and sequence-based scores and quality measures and with intuitive visualizations such that agreements and differences between alignments are easy to grasp. The user can thus make educated decisions about the structural similarity of two proteins and, if necessary, post-process alignments by hand. Furthermore, a comparative analysis allows to differentiate between proteins with one clear-case alignment on which various scorings agree and proteins with ambiguous alignments for which it depends on the application which alignment is preferable.

\section{Materials and methods}

\subsection{Structural alignment algorithm}

The exact algorithm used in CSA is based on an integer linear programming (ILP) model of the structural alignment problem as described in \cite{Wohlers2011}. Solutions to the ILP are generated using the approach from \cite{Andonov2011}. The algorithm combines branch-and-bound and Lagrangian relaxation, which can be seen as an iterative double dynamic programming method. The mathematical model supports a generic scoring scheme with positive and negative structural scores, sequence scores and affine gap costs. Many different scoring functions are special cases of this general scheme. Currently, CSA supports \dali~\cite{Holm1993}, \cmo \cite{Godzik1994}, \matras~\cite{Kawabata2000}, and \paul~\cite{Wohlers2010}.

\subsection{Webserver implementation}

The architecture of the web server is divided in a processing layer that computes  (C++) and evaluates (Python) alignments and an output layer, which generates W3C-validated HTML websites, interacts with the user and displays all information (PHP and Javascript). The interface between the two layers is a MySQL database.

The alignment engine for all our four currently supported scoring schemes is identical and implemented in C++ as a stand-alone program. User-adjustable parameters are the time limit of the computation, the maximum number of branch-and-bound nodes, and the number of Lagrangian iterations in each node. Furthermore, each scoring scheme has different parameters, for example, the use of C$_\alpha$ or C$_\beta$ inter-residue distances.

Computed or user-uploaded (in FASTA format) alignments are read into a Python class and subsequently written to the MySQL data base.
A second Python class handles the computation of different scores. It obtains the required structural information from the PDB files with the help of the Biopython package Bio.PDB~\cite{Hamelryck2003}. Tasks related to superpositioning are also handled by this package. Visualizations of distance and distance difference matrices are generated using the Python Imaging Library.

The website functions have been implemented in separate modules, which makes it easy to integrate additional structural alignment methods. The modularity is illustrated by the use of a tab menu. All web server functions are extensively documented, which is denoted by a question mark next to the respective section titles or table headers. Additionally, a documentation puts instructions and explanations into context. Notably, we documented all structural alignment scorings that are used within CSA and we provide the corresponding formulas and references. In the output layer, structures and their superpositions are visualized in Jmol (\url{http://www.jmol.org}) and images are generated using the PHP package pChart (\url{http://www.pchart.net/}).

\section{Case studies}

We illustrate the functionality of CSA using two case studies which are accessible from its main page via the links ``Example 1'' and ``Example 2''.

\subsection{Benefits of visualization and comparison}

The first case study deals with two proteins from the SISY data set~\cite{Mayr2007,Berbalk2009}, ubiquitin-binding protein CUE2 (PDB ID 1otr, chain A, 49 residues) and the CUE domain of activating signal cointegrator 1 complex subunit 2 (PDB ID 2di0, chain A, 71 residues). The proteins belong to the SISYPHUS~\cite{Andreeva2007} alignment AL00088995 of homologous proteins containing a CUE domain. The CUE domain is composed of a three helical bundle and it consists of 41 residues. It binds ubiquitin and is involved in protein degradation.

After specifying PDB IDs and chains on the main page of CSA, the user is redirected to the CSA evaluation environment. It is organized in tabs for the following tasks: overview on the protein structures, computing alignments using \cmo, \paul, \dali or \matras scoring, upload of external alignments, and the comparison of alignments.

The \emph{Structures} tab lists PDB IDs, PDB file names, selected chains and their lengths and amino acid sequences. Links to the PDB~\cite{Berman2000} and to iHOP~\cite{Hoffmann2005} are access points for additional information concerning the proteins and their function. Protein structures are visualized in Jmol. Their C$_\alpha$ and C$_\beta$ distance matrices and contact maps are visualized.

\begin{figure}[t]
\begin{center}
\includegraphics[width=12cm]{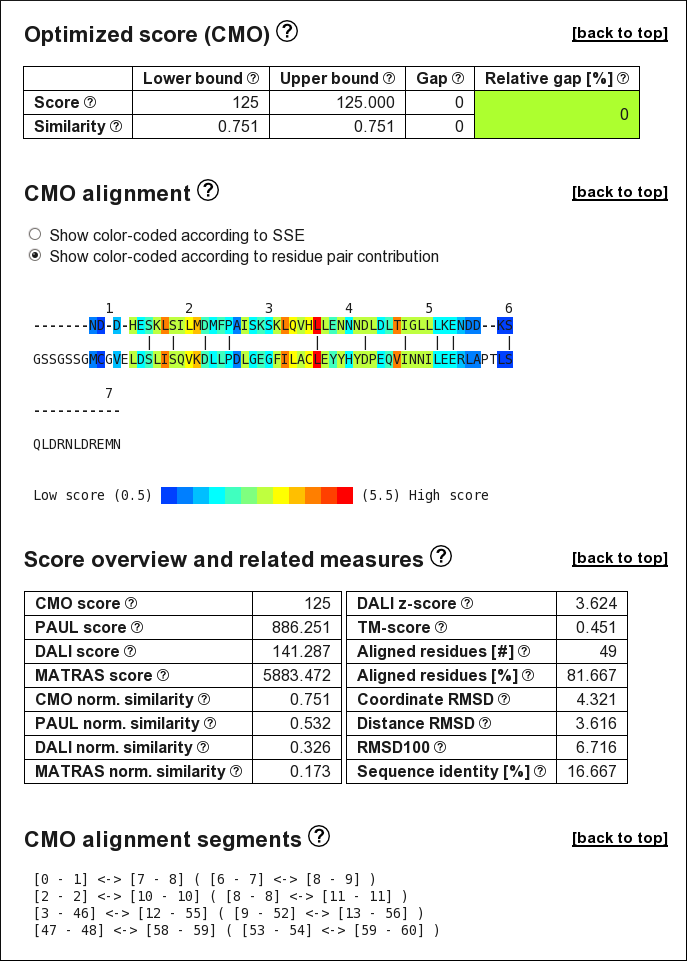}
\end{center}
\caption{Parts of the information displayed on the website for the \cmo alignment of 1otrA and 2di0A.}
\label{fig1}
\end{figure}

We compute \cmo, \paul, \dalix and \matrasx alignments using the default options, \ie, with a time limit of 30 CPU s. The setup of all four result pages is identical. Exemplary, we consider the CMO alignment page; parts of it are displayed in Figure \ref{fig1}.

\paragraph{Bounds on alignment score and similarity.}
The section \emph{Optimized score} lists the resulting scores: the raw score $s(A,B)$ of proteins $A$ and $B$ (here, the number of common contacts), and a similarity score that normalizes the raw score with respect to the self-similarity of the two proteins computed as 
$\frac{2s(A,B)}{s(A,A)+s(B,B)}$. Our exact algorithm returns lower an upper bounds ($\mathit{LB}$ and $\mathit{UB}$) on the raw and similarity scores. Based on these bounds, the relative gap in percent, $100\cdot\frac{\mathit{UB}-\mathit{LB}}{\mathit{LB}}$, quantifies by how many percent an alignment can at most be improved. Such a quality guarantee helps to quickly determine the progress of the computation as well as the similarity of the two proteins. If two proteins are dissimilar, the relative gap tends to be large, but the upper bound on the similarity score tends to be low from the beginning on. Aligning 1otrA with 2di0A w.r.t.\ \cmo yields 125 common contacts, and the corresponding similarity score on a scale from 0 to 1 is 0.751. The relative gap is 0\%, indicating that the top-scoring alignment has been found.

\paragraph{Structural conservation and variation.}
The \emph{Alignment} section displays the computed structural alignment. Residues are color-coded according to either SSE (helix, sheet, coil) as assigned by DSSP~\cite{Kabsch1983} or to residue pair score contribution. The second color-coding denotes how well single residue pairs are structurally conserved given the current alignment, \cf Fig.~\ref{fig1}. For the two proteins containing the CUE domain, this indicates that the first identically aligned leucines are structurally conserved, and in fact this position is part of a motif for binding ubiquitin that consists of an invariant proline and two highly conserved leucines~\cite{Shih2003}. Pairs of aligned residues with low score contribution highlight structural variations. In the \cmo alignment, the N- and C-terminal regions are little structurally conserved, as well as the residues in the region of the invariant proline within the CUE domain, because the proline is located in a turn. Such a visualization of residue score contribution can hint towards a manual modification of the alignment by removing aligned residues with low score. In fact, this is what happens in the top-scoring \dalix alignment of 1otrA and 2di0A, in which the four C-terminal residues with low \cmo score are excluded from the alignment.

\paragraph{Comprehensive alignment-related data.}
Additional to the alignment, CSA displays the aligned segments, both using sequential and PDB residue numbering, \cf Fig.~\ref{fig1}. Numerous raw alignment- and similarity scores are listed, for example the number of aligned residues, sequence identity and root mean square deviation (RMSD). Furthermore, some statistics concerning the alignment computation are given. These are the number of residues and inter-residue distances considered during computation. They greatly influence the memory consumption of the algorithm: the more inter-residue distances are considered, the more memory is needed and typically the larger the running time. Using default values, \cmo only considers distances smaller than 7.5 \AA, \paul considers distances smaller than 8.5 \AA\ (for C$_\alpha$ distances, 9.5 \AA), \matras uses distances up to 50 \AA\ and \dali all distances. Because server memory is shared among users, we currently restrict computations using the \dali or \matras scoring to protein pairs with average length less than 150 residues. The allocation time for setting up all data structures is given, as well as the time actually spent on computing the alignment. The number of visited branch-and-bound nodes gives a good estimate on the progress of the computation. The proteins are superposed according to the alignment and visualized in Jmol. The trace of aligned residues and the distance difference matrix is plotted.
 
We upload an additional alignment in the tab for the first custom alignment. This alignment aligns only the 38 residues that belong to the respective CUE domain and that are structurally equivalent according to SISY. Furthermore, we upload a second custom alignment which has been generated by the \dali server~\cite{Holm2010}. The \dali server uses a heuristic algorithm to find a good alignment according to the \dali score.

\begin{figure}[t]
\begin{center}
\includegraphics[width=8cm]{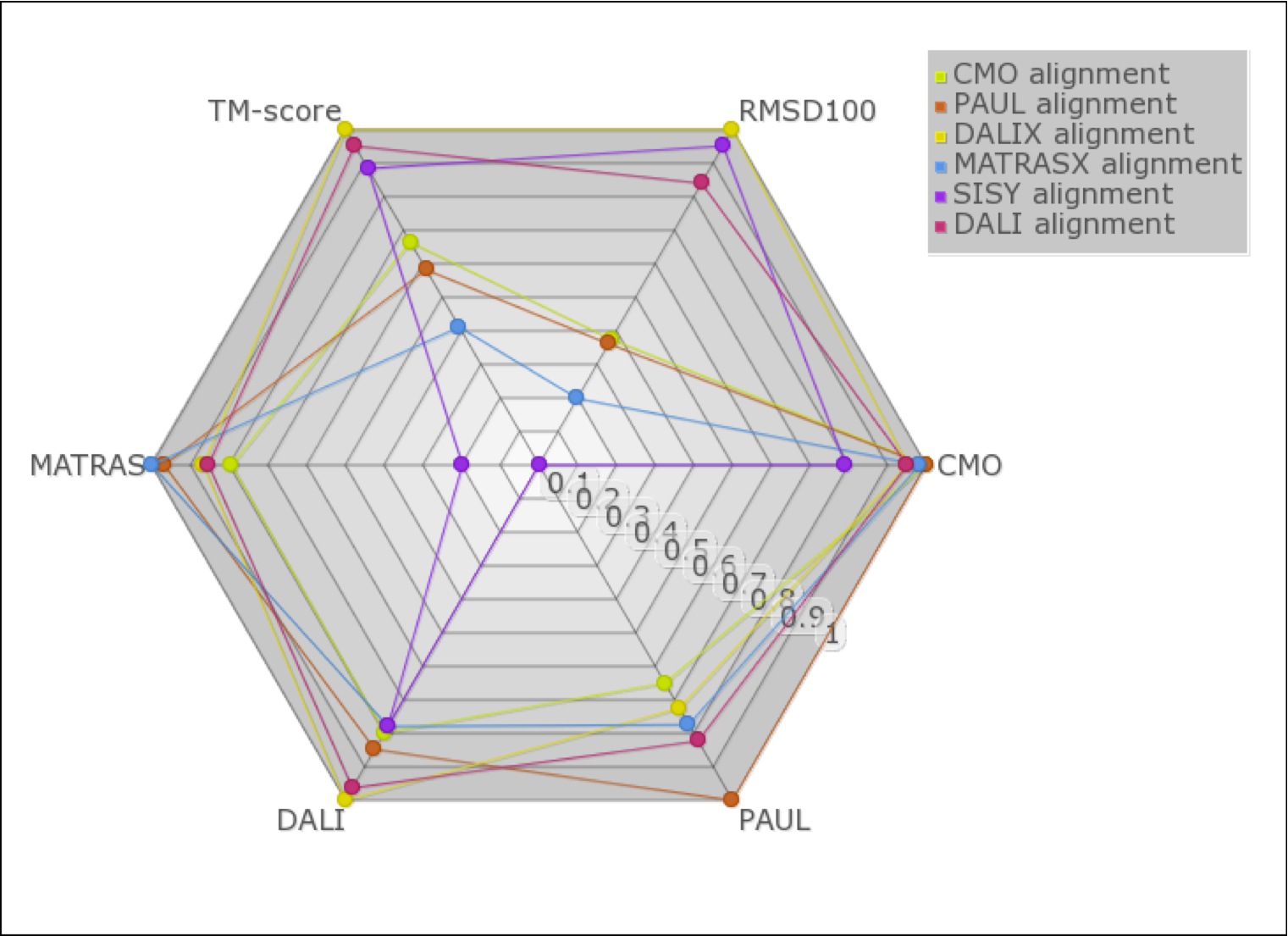}
\end{center}
\caption{A radar chart for comparison of alignment scores for six different alignments of 1otrA and 2di0A. The closer a point is to 1, the better the corresponding score. \cmo, \paul, \dalix and \matrasx alignments have been computed by our exact algorithm and are provably optimal concerning their respective score. The SISY reference alignment aligns 38 residues of the CUE2 domain. The \dali alignment was computed by the \dali server and has slightly lower \dali score than the optimal \dalix alignment. The reference alignment is far behind in all scores except RMSD100 and TM-score, for which it performs quite well. The \matrasx alignment performs especially poor for these two measures. Intuitively, the \dalix alignment is most preferable since it has optimal \dali and close to optimal \cmo, \paul and \matras scores, as well as the best TM-score and RMSD100.}
\label{fig2}
\end{figure}

\paragraph{Improving, verifying optimality and assessing quality of heuristic alignments.}
Many different scorings and quality measures can be compared in the \emph{Comparison tab}: the \cmo, \paul, \dali, and \matras raw and similarity scores, \dali z-score~\cite{Holm1998}, TM-score~\cite{Zhang2004}, number and percentage of aligned residues, coordinate and distance RMSD, RMSD100~\cite{Carugo2001}, and sequence identity. For 1otrA and 2di0A, all six computed and uploaded alignments differ from each other. While \cmo and \paul alignment were computed to optimality in less than a second, the \dalix alignment has the potential to be improved by up to 12\% and the \matrasx alignment by up to 24\%. We also observe that the alignment that was computed by the \dali server and then uploaded is better with respect to \dali score than the alignment computed by our exact algorithm within 30s. We thus increase the maximum running time for \dalix and \matrasx to 10 minutes. Now, both alignments are computed to provable optimality and our top-scoring \dalix alignment slightly improves the heuristic solution returned by the \dali server. \dalix and \matrasx alignments thus can be used to obtain quality guarantees for \dali or \matras alignments and in some cases also to either proof their optimality or to compute a better alignment.

\paragraph{Multi-criteria comparison and selecting a sound align\-ment.}
Alignment trace comparison as introduced in \cite{Godzik1996} gives a visual overview about agreements and differences between alignments. Here, any subset of alignments can be shown. Using this visualization, we find that all alignments (except the SISY reference, which excludes 3 residues in the center of the domain) correctly align all 41 residues of the CUE2 domain, and that they differ in aligning the neighboring N- and C-terminal residues. A radar chart compares the different scores, \cf~Fig.~\ref{fig2}. This chart helps to quantify score differences and allows to decide whether one alignment is clearly preferable, \ie, better with respect to all criteria. The chart also allows to make an intuitive decision which alignment is most appropriate in cases in which different scorings disagree as it is the case for 1otrA and 2di0A. Here, intuitively the \dalix alignment is the best choice since it performs good or best according to all criteria. 

Two residue pair lists show aligned residues that appear in all, resp.\ in the majority, of the alignments. They each constitute a consensus alignment. In the case of aligning 1otrA and 2di0A, we see that such a consensus is useful: all alignments only agree in aligning the CUE2 domain. The consensus thus highlights the structurally conserved and biologically relevant region of the alignment.

\subsection{Alignment of flexible proteins}
 
We illustrate the usefulness of comparing structural align\-ments in the case of protein flexibility. This is a challenge for most structural alignment methods because flex\-ible proteins typically do not superpose well unless the flexibility is accommodated for, \eg, by explicitly introducing a hinge. 

\begin{figure}[t]
\begin{center}
\includegraphics[width=3cm]{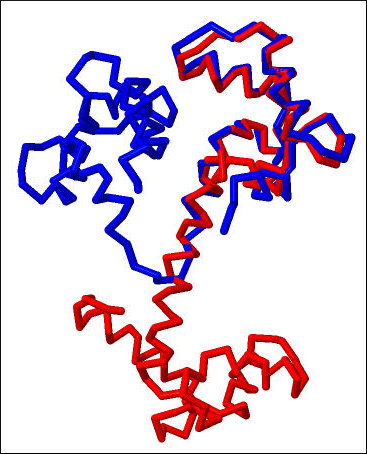}
\hspace{0.01cm}
\includegraphics[width=4cm]{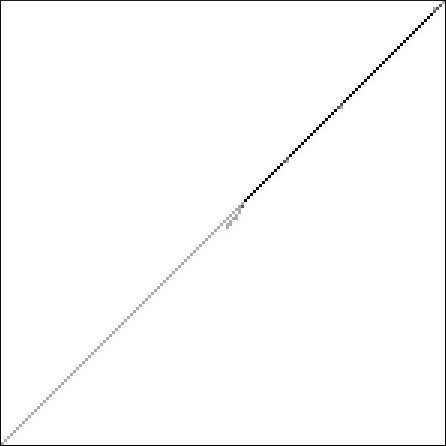} \\
\vspace{0.1cm}
\includegraphics[width=8cm]{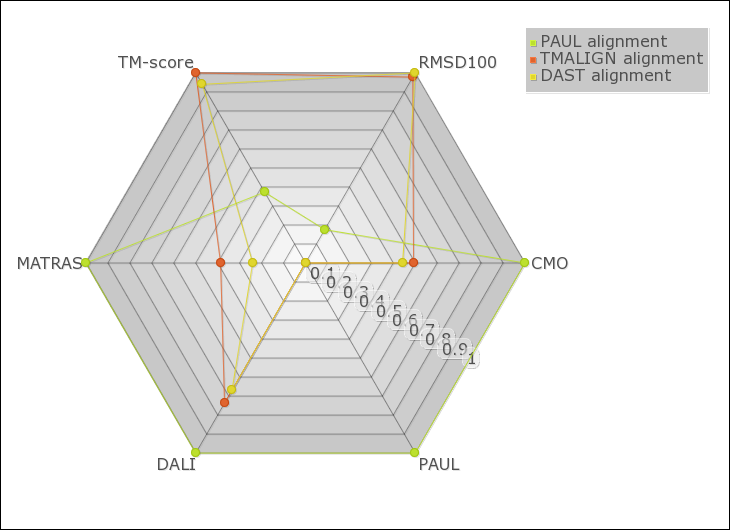} \\
\vspace{0.1cm}
\includegraphics[width=12cm]{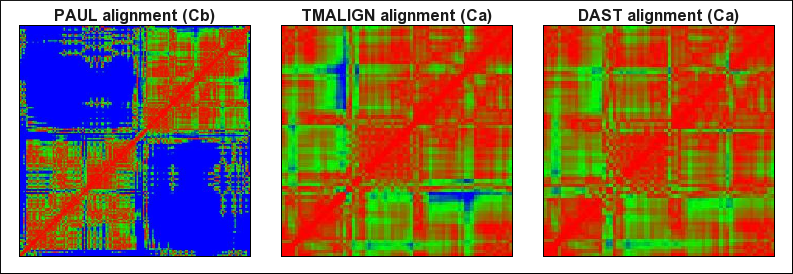}
\end{center}
\caption{Top left: The two calmodulin conformers (PDB IDs 4cln and 2bbm) superpositioned according to the TM-alignment, which aligns only one of the two regions that move relative to each other. Top right: Comparison of the alignment traces. Each axis corresponds to one conformer. Black boxes denote residue pairs aligned by all three scorings, \paul, \tmalign and \dast. Light gray denotes residue pairs aligned by only one scoring. An intermediate shade of gray denotes agreement of two scorings. \paul aligns all residues of the two conformers, \tmalign and \dast the C-terminal region. Center: The radar chart illustrates the difference between scorings that are more in favor of a flexible alignment, \ie \cmo, \paul, \dali and \matras, and scorings that are more in favor of a rigid superposition of low RMSD. Bottom: The distance difference matrices illustrate the difference between the flexible \paul alignment, that aligns all residues in spite of large distance differences (colored blue), and the TM- and \dast alignment, which exclude large distance differences, but only align the C-terminal region. }
\label{fig3}
\end{figure}

\paragraph{Comparing flexible and rigid scoring schemes.}
We align two conformations of the calmodulin protein (PDB IDs 4cln, chain A and 2bbm, chain A, with a length of 148 residues). In structure 4clnA, calmodulin is bound to a ligand, in 2bbmA it is unbound. In bound conformation, a central helix is split and the components at the ends of the helix are moved towards each other. We align the two conformations using \cmo and \paul. We furthermore upload the alignments computed by \tmalign~\cite{Zhang2005}, an algorithm maximizing the TM-score, and \dast~\cite{MalodDognin2010}, a local structural alignment method that determines the longest alignment with distance RMSD less than 4 \AA. We find that both \cmo and \paul correctly align the two conformations over their entire length. To keep the comparison concise, we thus exclude the \cmo alignment from the following visualizations. Figure~\ref{fig3} displays the  two conformers superposed according to the TM-alignment and the alignment trace comparison. While \paul aligns all residues of the two conformers correctly, \tmalign aligns only the C-terminal, rigidly superposable region (except the C-terminal residue). \dast also aligns the C-terminal region, but excludes and shifts further residues from the alignment. The radar chart comparing the different scores as well as the distance difference matrices displayed in Figure~\ref{fig3} show why: while \cmo, \paul, \dali and \matras scoring by far favor the alignment of the entire conformers, TM-score as well as RMSD100 clearly favor the TM- and \dast alignment, which has a much smaller RMSD, but aligns only the C-terminal region.

\paragraph{Detecting flexibility and hinges.}
For each alignment we display the distance difference matrix. This is a symmetric square matrix with entries $|d_{ij}^A-d_{ij}^B|$ at position $(i,j)$, where $i$ is the $i$-th aligned position and $j$ the $j$-th aligned position. Here, distance differences are visualized using a color gradient in which 0 \AA\ is colored red, 2.5 \AA\ green 5 \AA\ blue. Regions with low inter-residue distance differences correspond to rigidly superposable fragments. For the \paul alignment of 4clnA and 2bbmA, red blocks in the distance difference matrix indicate that both the N-terminal and C-terminal regions can be superpositioned very precisely. The distance differences between these two regions, however, are large, denoted by the blocks in blue color. The two regions can thus only be well superpositioned individually. A hinge is present at the residue bordering the two blocks (position 80)~\cite{Emekli2008}. \tmalign and \dast align only the C-terminal region, thus avoiding any large distance differences. \dast is more restrictive in excluding large distance differences, it does not align a few residues that are still aligned by the TM-alignment and which have distance differences larger than 5 \AA, colored in blue.

Scores as \cmo and \paul, which implicitly ignore RMSD, are useful to gain information about flexible regions. While this feature is beneficial for flexible proteins it may also introduce flexibility where this is not appropriate. Protein similarities consisting in compact, well superposable fragments are therefore often better detected by maximizing scores like the TM- or the \dast score.

\section{Conclusion}

Different structural alignment scoring functions have different strengths and weaknesses. Which scoring to use depends on the application and on the structural relationship of the investigated proteins. Their different focus on handling various aspects of structural similarity is one reason why there are many different structural alignment scorings and programs and no consensus which combination is best.

We therefore consider it beneficial to compute alignments using different scoring schemes and algorithms and to compare them in order to gain insight into their structural relationship. The CSA web server provides the tools for such a comparison. CSA allows to compute alignments with various scorings, returns a quality guarantee for the alignments and enables the user to additionally evaluate and compare uploaded alignments. In the most common case in which scorings and alignments disagree, it facilitates evaluating the agreement and differences between them and selecting the most suitable alignment.

\section{Funding}

This work was partly supported by the DFG [KL 1390/2-1].

\section{Acknowledgments}
We thank Maarten Dijkema for his help with setting up the server. Further, we thank Mathilde Le Boudic-Jamin for extensively testing CSA and Thomas Metzler for his idea of using a radar chart for comparison.

\subsubsection{Conflict of interest statement.} None declared.

\bibliographystyle{plain}
\bibliography{biblio}

\end{document}